\documentclass{elsart3}
\usepackage[xdvi]{graphicx}
\usepackage{amssymb}
\journal{Nuclear Instruments and Methods in Physics Research, section A}

\begin{document}
\begin{frontmatter}

	\title{Development of an advanced Compton camera with gaseous TPC and scintillator}
	\author{A. Takada\corauthref{cr}}
	\corauth[cr]{Tel: +81-75-753-3869; Fax: +81-75-753-3799}
	\ead{takada@cr.scphys.kyoto-u.ac.jp}
	\author{, K.Hattori, H.Kubo, K.Miuchi, T. Nagayoshi, H.Nishimura}
	\author{Y. Okada, R. Orito, H. Sekiya, A. Takeda, T. Tanimori}
	\address{Cosmic-Ray group, Department of Physics, Graduate School of Science, Kyoto University}

	\begin{abstract}
		% Text of abstract
		A prototype of the MeV gamma-ray imaging camera based 
		on the full reconstruction of the Compton process has been developed.
		This camera consists of a micro-TPC 
		that is a gaseous Time Projection Chamber (TPC)
		and scintillation cameras.
		With the information of the recoil electrons and the scattered gamma-rays,
		this camera detects the energy and incident direction of each incident gamma-ray.
		We developed a prototype of the MeV gamma-ray camera 
		with a micro-TPC and a NaI(Tl) scintillator,
		and succeeded in reconstructing the gamma-rays from 0.3 MeV to 1.3 MeV.
		Measured angular resolutions of
		ARM (Angular Resolution Measure) and SPD (Scatter Plane Deviation)
		for 356 keV gamma-rays were $18^\circ$ and $35^\circ$, respectively.
	\end{abstract}


\end{frontmatter}

% main text
\section{Advanced Compton imaging}
In spite of the significance of the MeV gamma-ray imaging in Astronomy and Medical imaging,
the quality of MeV gamma-ray images obtained with existing cameras,
such as a position-sensitive detector with a collimator \cite{kurfess}
and classical Compton imaging \cite{schonfelder},
are not adequate yet.
The deterioration of the images is attributed to the large background
made by the scatterings in the collimator and other materials nearby
and the ghost images intrinsic to the classical Compton method.
Therefore we need a new method 
which can reconstruct the incident direction completely for a single photon
and can reject the background in order to obtain clearer MeV and sub-MeV gamma-ray images.

\begin{figure}
	\begin{center}
		\includegraphics[width=5.1cm]{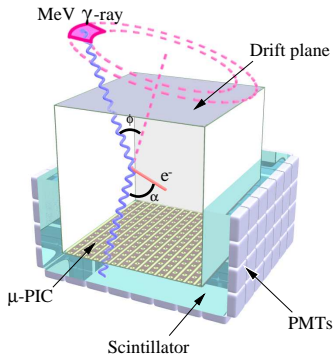}
		\caption{A schematic view of our detector.}
		\label{fig:detector}
	\end{center}
\end{figure}

We are developing a detector based on a new method: 
the advanced Compton imaging \cite{tanimori04}.
Fig. \ref{fig:detector} shows the schematic view of our detector.
Our detector consists of a micro-TPC, which is a three dimensional tracker of charged particles, 
and a surrounding position-sensitive scintillator.
When the incident gamma-rays Compton-scatter in the micro-TPC \cite{miuchi03},
the recoil electrons are detected by the micro-TPC and 
the scattered gamma-rays are absorbed in the scintillator.
The difference with classical Compton imaging is the 3D tracking of the recoil electrons.
With the information of the recoil electron and the scattered gamma-ray,
the energy and direction of the incident gamma-rays can be reconstructed for a single photon.
Besides, the residual angle $\alpha$ 
(the angle between the scattering direction and recoil direction, 
as shown in Fig. \ref{fig:detector}), 
provides a kinematical background rejection.
The angle $\alpha$ can be not only measured geometrically ($\alpha_{\rm geo}$) 
but also calculated by the kinematics of Compton scattering ($\alpha_{kin}$).
We can reject the backgrounds and obtain high quality gamma-ray images
by requiring $\alpha_{geo} \simeq \alpha_{kin}$.
The Medium Energy Gamma-ray Astronomy telescope (MEGA) \cite{zoglauer} is also a detector
based on advanced Compton imaging,
whose tracker is a stack of silicon strip detectors.
The recoil electrons multiple scatter in the solid tracker,
and initial directions are easily lost.
For reducing the multiple scattering,
we need to use low Z and low density material for the tracking detector,
and detect the recoil direction 
with the initial short part ($\sim$ a few mm) of the electron track.
Therefore, we adopt the micro-TPC, which is a gaseous tracker with a good position resolution.
The micro-TPC provides an accurate measurement of the recoil direction, 
and allows the sensitivity to the sub-MeV and MeV gamma-rays.

\section{Performance of the prototype detector}
\subsection{setup}
We developed a prototype of the advanced Compton camera for conceptual measurement.
The prototype detector consists of a micro-TPC of $10\times 10\times 8$ cm$^3$ 
(fiducial volume: $8\times 8\times 7.5$ cm$^3$) and
a scintillation camera of $10\times 10\times 2.5$ cm$^3$ 
(fiducial volume: $7\times 7\times 2.5$ cm$^3$).
The scintillator is a monolithic NaI(Tl) with an Anger type photo-readout.
The photo-readout consists of 5$\times$5 single anode PMTs of 3/4 inch diameter. 
We get the hit position in the scintillator by the center of the photo-distribution,
and we used the center of the thickness of the scintillator as the depth of the interaction.
For 356 keV gamma-rays,
the energy resolution of the scintillator is 9.5\% (FWHM),
and the position resolution is 4.5 mm (RMS).

The micro-TPC is a time projection chamber with a $\mu$-PIC \cite{nagayoshi}.
The $\mu$-PIC, which is a gaseous 2D imaging detector 
with the micro pixel electrode made by the Print Circuit Board technology,
has a good gas gain uniformity (4.5\% RMS) and fine position resolution ($\sim 120 \mu$m RMS).
We filled the drift volume with Ar+C$_2$H$_6$ (90:10) flow of 110 ccm
and electric field of 400 V/cm.
We used the $\mu$-PIC (TOSHIBA SN040426-1) as a readout with a gas gain of $\sim$5000.

We placed a gamma-ray source at $\sim$30 cm from the micro-TPC,
and obtained the images of $^{137}$Cs (662 keV, 0.93 MBq), $^{54}$Mn (835 keV, 0.61 MBq),
$^{133}$Ba (356 keV, 1.1 MBq) and $^{60}$Co (1173 keV and 1333 keV, 0.84 MBq).
For the details of the setup, see Refs. \cite{tanimori04} and \cite{takeda}.

\subsection{analysis}
The track of the recoil electron, the energy and the direction of scattered gamma-ray
are measured for the reconstruction of the Compton scattering.
We did not use the energy of the recoil electron,
because the volume of the prototype micro-TPC is so small 
that the recoil electrons of more than 100 keV escape from the micro-TPC.
Therefore we used the energy of an incident gamma-ray 
as a known parameter for the reconstruction.

\begin{figure}
	\begin{center}
		\includegraphics[width=5.6cm]{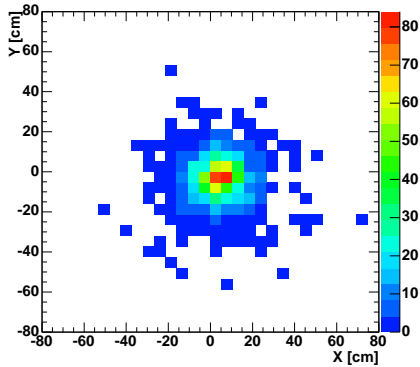}
		\caption{The image of $^{133}$Ba. The source position is $(12$ cm$, -4.0$ cm$)$.
			In this image, the white region means 0 events.}
		\label{fig:image}
	\end{center}
\end{figure}

Now we select the events under the following conditions.
\begin{itemize}
	\item fiducial volume:
		We use the events which are in the fiducial volume of the micro-TPC and the scintillator.
	\item $N_{hit} \geq 3$:
		To know the recoil direction,
		we fit the initial 3 points of the electron track with a straight line.
		Therefore we required number of hits ($N_{hit}$) $\geq 3$.
	\item forward-coming gamma-rays:
		We obtain the gamma-ray images only for the forward-coming gamma-rays,
		and exclude backward-coming gamma-rays.
	\item multiple scattering:
		The tracks of bad fitting (RMS $> 2$ mm) are not included in the images.
	\item limit of $\alpha$:
		We required $\alpha > \alpha_{lim}$,
		where $\alpha_{lim}$ is the kinematically minimum of the angle $\alpha$
		and calculated for a given energy of the incident gamma-ray.
	\item recoil angle:
		The recoil angle ($\alpha - \phi$ in Fig. \ref{fig:detector}) does not exceed $90^\circ$.
		We required $\alpha_{geo} - \phi < 90^\circ$.
	\item $\alpha$ cut (kinematical background rejection):
		For the kinematical fit,
		we required $\left| \alpha_{geo} - \alpha_{kin} \right| \leq 7.5^\circ$
		as the kinematical constraint.
\end{itemize}
With these conditions, 
we obtained images of gamma-rays for 0.3-1.3 MeV with a high signal-to-noise ratio.
Fig. \ref{fig:image} is the image for 356 keV gamma-rays.
In this figure the white region means 0 events,
and it shows the signal-to-noise ratio is very high.
It should be noted that we can obtain images of low energy gamma-rays
because of the gaseous tracker.

\subsection{results}
\begin{figure}	
	\begin{center}
		\includegraphics[width=\linewidth]{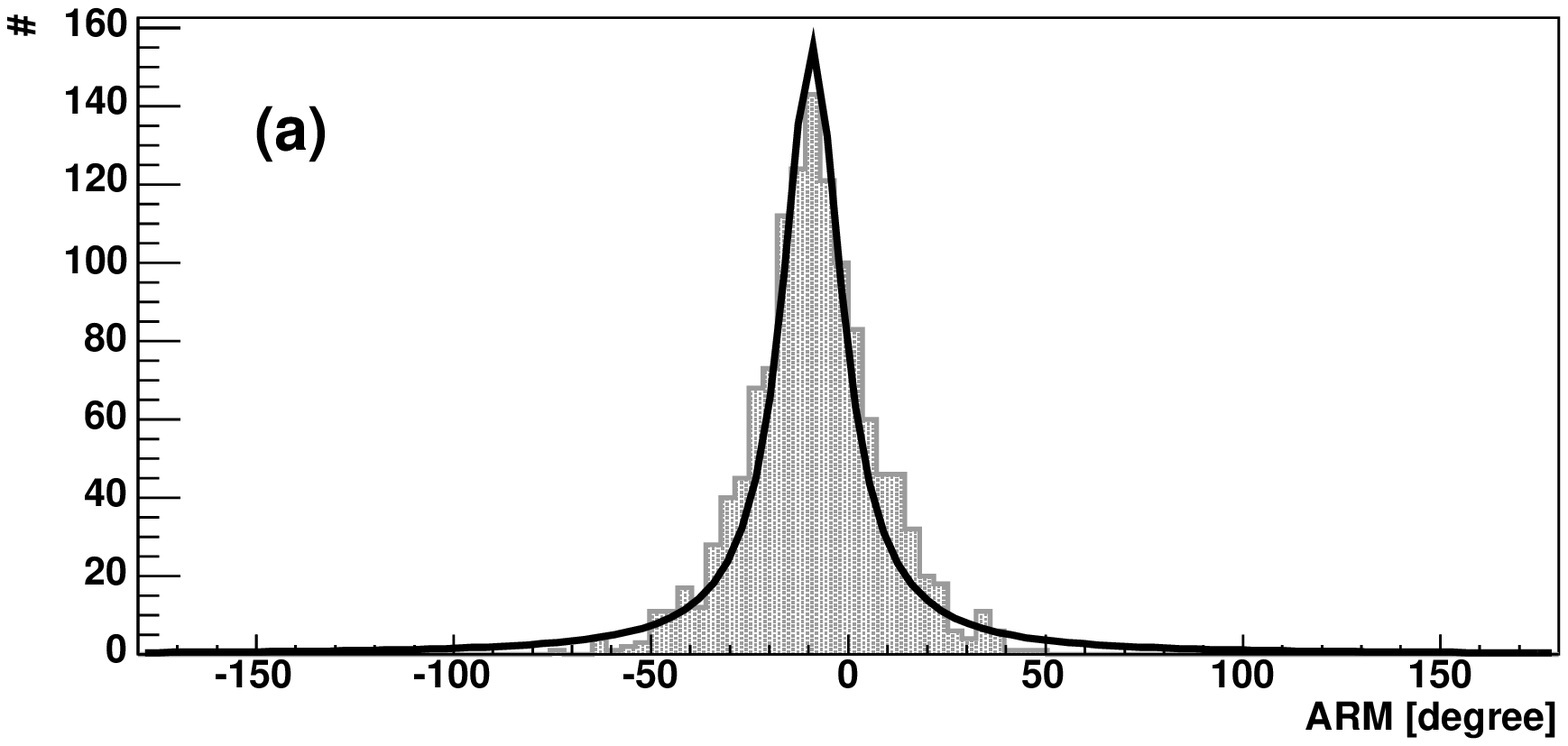}
		\includegraphics[width=\linewidth]{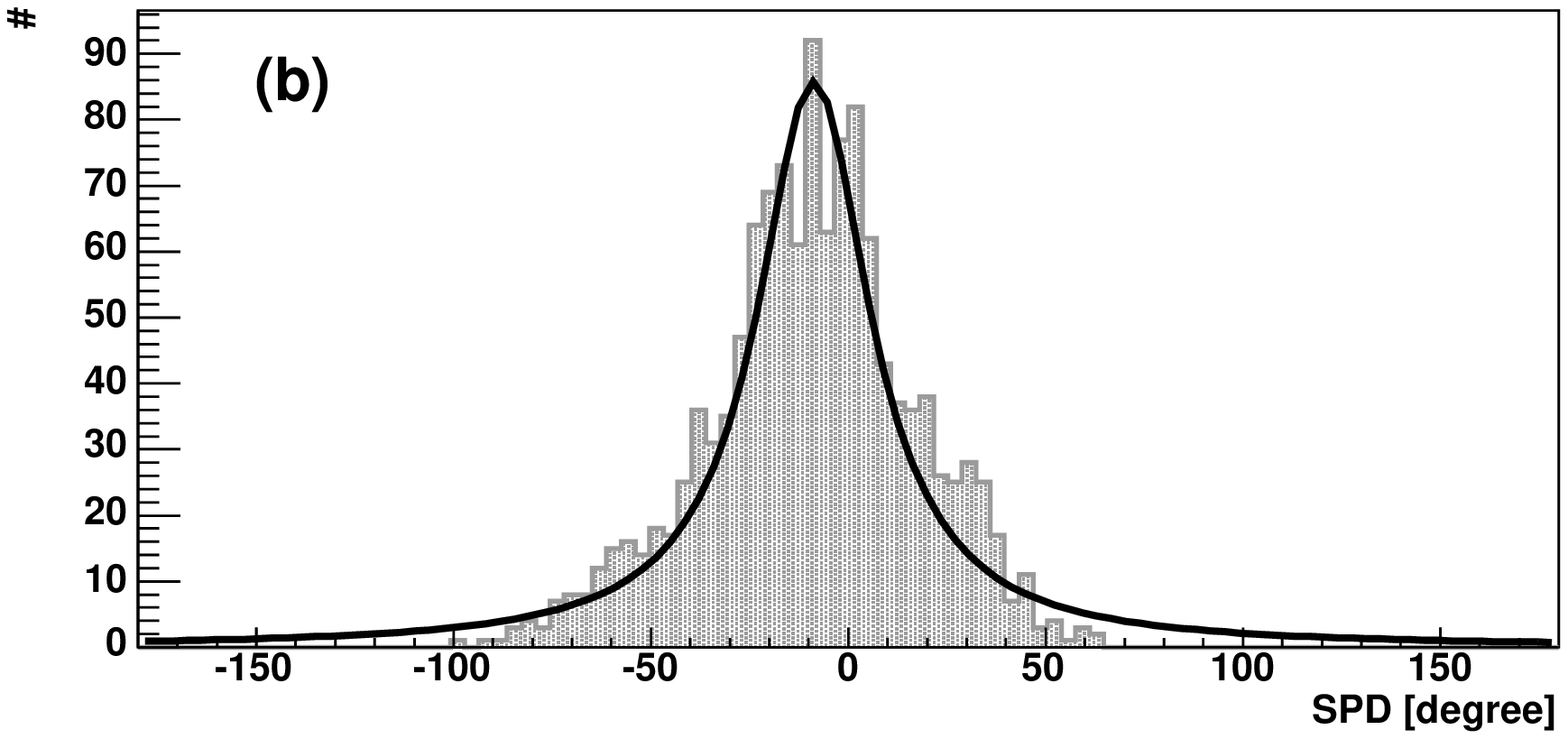}
		\caption{The distributions of ARM (a) and SPD (b) for $356$ keV gamma-rays.
			The solid lines are best-fit Lorentzians.}
		\label{fig:distribution}
	\end{center}
\end{figure}
The angular resolution of an advanced Compton camera is defined by two angle parameters,
and the error region is a sector form.
One is the Angular Resolution Measure (ARM),
which is defined as the angle between detected scattering angle and real scattering angle.
The other is the Scatter Plane Deviation (SPD),
which means the determination accuracy of the scatter plane.
Fig. \ref{fig:distribution} is the distribution of ARM and SPD for 356 keV gamma-rays.
In each figure, the solid line is the best-fit with a Lorentzian.
For 356 keV gamma-rays, the FWHM resolutions of ARM and SPD
are $18^\circ$ and $35^\circ$, respectively.
It should be noted that there is no background and no floor in either distribution.

\begin{figure}
	\begin{center}
		\includegraphics[width=\linewidth]{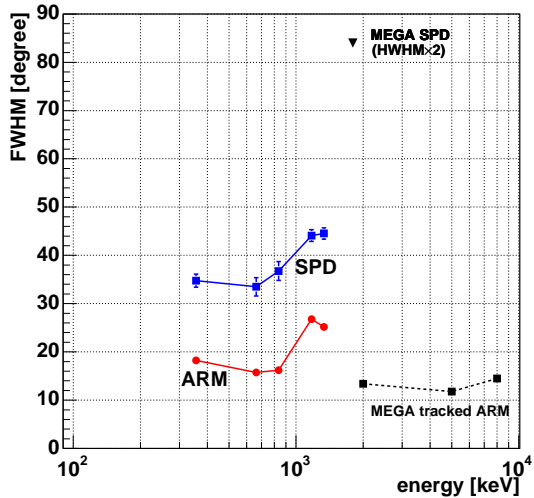}
		\caption{ARM and SPD of the prototype (FWHM). 
			For comparison, also shown are the ARM and SPD of MEGA detector \cite{zoglauer}.}
		\label{fig:arm_spd}
	\end{center}
\end{figure}

The energy dependences of ARM and SPD are shown in Fig. \ref{fig:arm_spd}.
Because the tracking efficiency is not high enough yet, 
ARM and SPD are not very good above 1 MeV.
The ARM and SPD of MEGA\cite{zoglauer} are also shown in Fig. \ref{fig:arm_spd} for comparison.
It is clearly shown that our detector has better SPD resolution
and is sensitive even to the gamma-rays below 511 keV due to the gaseous tracker.

\begin{figure}
	\begin{center}
		\includegraphics[width=\linewidth]{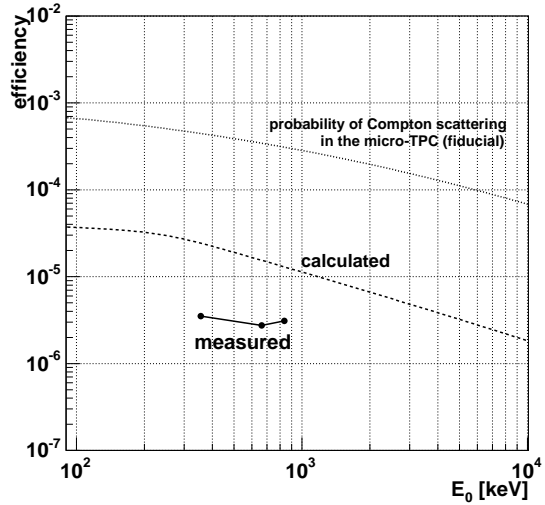}
		\caption{The detection efficiency of the prototype (solid line).
			The dotted line is the probability of Compton scattering in the micro-TPC,
			the dashed line is the probability of 
			Compton scattering in the micro-TPC and interaction in the prototype scintillator.}
		\label{fig:efficiency}
	\end{center}
\end{figure}

The detection efficiency of the prototype is shown in Fig. \ref{fig:efficiency}.
For 356 keV gamma-rays, the measured detection efficiency is $3.5\times 10^{-6}$.
The calculated efficiency of this experimental setup is $2.4\times 10^{-5}$.
The difference between the measurement and calculation is attributed to the tracking inefficiency,
which we think will improve by the improvement of the $\mu$-PIC.
By comparison with the probability of Compton scattering in the micro-TPC,
the calculated efficiency is very small,
because the area covered by the scintillator is not large enough.
The calculated efficiency is expected to be very close to the Compton scattering probability
when we develop a larger scintillator.

Now, for the higher efficiency and better angular resolutions,
we are developing a large size micro-TPC with a $30\times 30$ cm$^2$ $\mu$-PIC
and the pixel-type scintillation cameras with larger areas, as shown in Fig. \ref{fig:detector}.
When we develop a micro-TPC of large size,
this camera will have the sensitivity to higher energy gamma-rays.

\section{Conclusion}
We developed a prototype advanced Compton camera,
and reconstructed the Compton-scattering of gamma-rays photon by photon.
This result proved the principle of the gamma-ray detection.
Because we use a gaseous tracker with a good position resolution,
our detector can detect lower energy gamma-rays and has a good SPD resolution.
That clearly shows the significance of the recoil electron tracks for the Compton imaging.

\section*{Acknowledgment}
This work is supported by the 
Grant-in-Aid 
from the Ministry of Education, Culture, Sports, Science and Technology (MEXT) of Japan
and Grant-in-Aid for the 21st Century COE ``Center for Diversity and Universality in Physics'' 
from MEXT.

\end{document}